\documentclass[12pt]{article}

\usepackage[a4paper,left=1in,top=1in,width=6.3in,height=9in]{geometry}
\usepackage{graphicx} 
\usepackage{bm}
\usepackage{tabularx}
\usepackage{authblk}
\usepackage[dvipsnames]{xcolor}
\usepackage{amsmath}
\usepackage{amsthm}
\usepackage{pdflscape}
\usepackage{mathtools}
\usepackage{amssymb}
\usepackage[title]{appendix}
\usepackage{multirow}
\usepackage[sort]{natbib}
\usepackage{subcaption}

\usepackage{wrapfig} 
\usepackage{tikz}

\usetikzlibrary{arrows.meta, positioning, shapes.geometric}
\usepackage{pgfplots}
\pgfplotsset{compat=1.18}
\usepgfplotslibrary{polar}

\newcommand{\mycomment}[1]{}

\numberwithin{equation}{section}

\usepackage[width=.75\textwidth]{caption} 
\numberwithin{equation}{section}
\usepackage[utf8]{inputenc}
\usepackage{url}
\usepackage{hyperref}
\usepackage{amsthm}

\newcolumntype{a}{>{\columncolor{red}}c}
\newcolumntype{g}{>{\columncolor{green}}c}

\newcommand{\nil}[1]{}

\newcommand{\dummy}[1]{}

\title{\bf A Cylindrical Galton Board  at the Galton Board's 150th Anniversary}
\author{ Kanti V. Mardia $^1$, Colin Goodall $^1$ and John Rubbo $^2$ }
\affil{$^1$ k.v.mardia@leeds.ac.uk; colin.goodall1@gmail.com; University  of Leeds, Leeds LS2 9JT, UK \\
$^2$ jrubbo48@aol.com; Little Brown House, Matawan NJ, USA  }

\date{} 
\begin{document}
\maketitle

\begin{abstract}
The celebrated Galton board remains a widely used device for demonstrating how
repeated Bernoulli trials on a triangular lattice generate an approximately
normal distribution. Marking the 150th anniversary of Galton’s 1875 construction,
this paper revisits the historical apparatus and extends it to a cylindrical
setting in which the peg lattice is wrapped around a cylinder, enforcing angular
periodicity and producing height‑dependent behaviour absent from the classical
design. This modification connects Galton’s original demonstration of variation
and the emergence of the normal distribution with modern developments in
circular statistics, providing a physical realisation of binomial random walks on
a circular–linear product space. We distinguish configurations where the wrapped
lattice subtends only a strict arc from those spanning the full circumference,
and show how the resulting geometry yields wrapped binomial and wrapped normal
behaviour. We describe the design and construction of our physical model, outline
practical considerations for replication, and analyse the statistical properties
and pedagogical applications of this cylindrical variant, which offers a modern
reinterpretation of Galton’s pioneering work.
\end{abstract}

 
\section{Introduction}
Modern implementations of the celebrated Galton board continue to be widely used.
For example, the STEM Galton Board produced by Index Fund Advisors (IFA) has sold
more than 60{,}000 units worldwide and is accompanied by extensive instructional
material, videos, and a detailed user guide linking the device to random
behaviour in financial markets. This contemporary version is described at \\
\url{https://www.ifa.com/galtonboard}. \\ The enduring popularity of such devices
reflects the continuing pedagogical value of physical demonstrations of
probabilistic variation

The classical planar Galton board provides a vivid illustration of how a normal
distribution emerges from repeated independent binary decisions. Balls traverse a
triangular array of pegs and accumulate in bins at the base, but the planar
geometry necessarily introduces artificial lateral boundaries that break
translational symmetry and influence the resulting distribution. These geometric
constraints are rarely discussed but have important implications for the
interpretation of the device.

In this paper, we propose a cylindrical Galton board obtained by wrapping the peg
lattice continuously around a solid cylinder. This construction enforces periodic
boundary conditions in the horizontal direction and yields a natural circular bin
structure. The resulting device provides a physical realisation of a binomial
random walk on the manifold $S^1 \times \mathbb{R}$, leading naturally to the
wrapped normal distribution. The cylindrical geometry restores horizontal
symmetry, eliminates boundary effects, and offers a new perspective on Galton’s
original pedagogical aims.

The remainder of the paper is organised as follows. Section~\ref{Sec: Hist}
summarises the historical development of the Galton board and its role in
nineteenth‑century statistical thought. Section~\ref{Sec: Formulation} presents a
detailed parametrisation of a Galton board whose peg lattice is embedded on the
surface of a solid cylinder. Section~\ref{Sec: wrapped normal} describes the
underlying wrapped distributions, including the wrapped binomial and the wrapped
normal. Section~\ref{sec:dynamics} discusses stochastic dynamics on the
cylindrical board. Section~\ref{Sec: Physical} reports on the construction of our
physical cylindrical board and its operation. Section~\ref{Sec: Disc} concludes
with a discussion. Appendix \ref{Sec: Appendix} constructs a cylindrical model by 
wrapping a synthetic Galton board.
\section{Historical Context}
\label{Sec: Hist}

Francis Galton’s probability machine, later known as the Galton board or
\emph{quincunx}, occupies a distinctive place in the development of
nineteenth‑century statistical thought. Although the device is now familiar from
modern demonstrations of the central limit theorem, its origins lie in Galton’s
broader programme of illustrating variation, error, and regression through
physical apparatus.

Galton first alluded to the mechanism in his 1875 paper \citep{Galton1875}, where
he used a peg‑based device to demonstrate regression to the mean and the
emergence of the normal distribution. A fuller description appeared in
\citet{Galton1877}, and an extended treatment was later included in his 1889 book
\citep[pp.~63--70]{Galton1889}, \emph{Natural Inheritance}. In these works, the
board served not merely as a curiosity but as a concrete embodiment of the
binomial law of error, illustrating how aggregate regularity arises from
individual randomness.

The apparatus also featured prominently in Galton’s public lectures, including
his Royal Society demonstrations, where it provided a striking visualisation of
the central limit phenomenon. Stigler’s historical analysis
\citep[pp.~275--281]{stigler1986} situates the device within the broader
nineteenth‑century effort to communicate probabilistic ideas to scientific and
lay audiences alike. The quincunx exemplified Galton’s belief that statistical
principles could be made accessible through carefully designed physical models.

Digital versions of the Galton board are now widely available, including
interactive apps and online simulations that allow users to vary the number of
rows, probabilities, and visual settings. These modern implementations continue
Galton’s original aim of making probabilistic ideas accessible to broad
audiences, and illustrate the enduring pedagogical appeal of the device.

For completeness, readers may wish to view a demonstration of Galton’s original 1873 apparatus, which can be seen in operation during the first two minutes of the following video:
\[
\text{https://www.youtube.com/watch?v=SZoDNfVFS7I}.
\]
These historical materials highlight the pedagogical and conceptual motivations
behind the device and provide context for our cylindrical extension, which seeks
to preserve Galton’s original aims while addressing geometric limitations
inherent in the planar design.

\section{Cylindrical Geometry and Peg Lattice}
\label{Sec: Formulation}

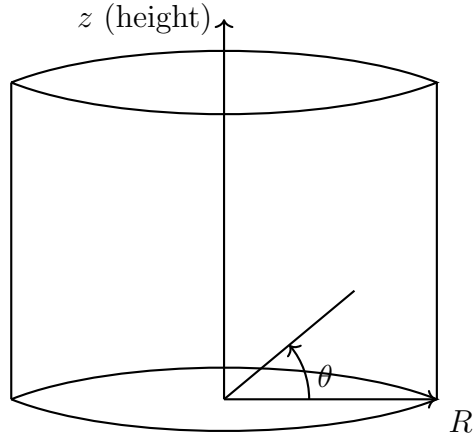
\begin{figure}[h!]
\centering
\begin{tikzpicture}[scale=1.4]

\draw[thick] (-2,3) .. controls (-1,3.4) and (1,3.4) .. (2,3)
             .. controls (1,2.6) and (-1,2.6) .. (-2,3);

\draw[thick] (-2,0) .. controls (-1,0.4) and (1,0.4) .. (2,0)
             .. controls (1,-0.4) and (-1,-0.4) .. (-2,0);

\draw[thick] (-2,3) -- (-2,0);
\draw[thick] (2,3) -- (2,0);

\draw[->,thick] (0,0) -- (0,3.6) node[left] {$z$ (height)};

\draw[thick,->] (0,0) -- (2,0) node[below right] {$R$};

\draw[thick] (0,0) -- ({1.6*cos(40)},{1.6*sin(40)});

\draw[->,thick] (0.8,0) arc[start angle=0, end angle=40, radius=0.8];

\node at (0.95,0.22) {$\theta$};

\end{tikzpicture}
\caption{Cylindrical Galton board with labelled axes: drops evolve vertically
($z$ direction), collisions rotate angular position ($\theta$), and bins wrap
around the circular base.}
\label{fig:cylinder}
\end{figure}

We introduce here the geometric and lattice parameters required for a cylindrical
Galton board. A right circular cylinder of radius $R$ is parameterised by angular
coordinate $\theta \in [0,2\pi)$ and vertical coordinate $z \in \mathbb{R}$, as
illustrated in Figure~\ref{fig:cylinder}. The circumference is $C = 2\pi R$.

We consider peg lattices on the cylinder where adjacent pegs on the same row are
separated by horizontal arc length $d>0$, and successive rows are separated
vertically by $h>0$. The angular spacing between neighbouring pegs is therefore
\[
\Delta\theta = \frac{d}{R},
\]
and we choose $d$ (equivalently $\Delta\theta$) so that the circle is divided
into an integer number $M$ of angular slots:
\[
M = \frac{C}{d} = \frac{2\pi}{\Delta\theta}.
\]

Rows are indexed by $i = 0,1,\dots,n$. Row $i$ contains $i+1$ pegs in the
unwrapped triangular lattice and
\[
i_M = \min(M,\, i+1)
\]
pegs when wrapped onto the cylinder. Table~\ref{tab:parameters1} summarises the
geometric and lattice parameters.

\begin{table}[h]
\centering
\caption{Summary of cylindrical Galton board parameters}
\label{tab:parameters1}
\begin{tabular}{ll}
\hline
\textbf{Parameter} & \textbf{Description} \\
\hline
$M$                 & Number of angular bins \\
$\Delta\theta$      & Angular spacing between pegs (bin width) \\
$d$                 & Horizontal arc length between adjacent pegs \\
$h$                 & Vertical spacing between rows of pegs \\
$n$                 & Number of peg rows \\
$H$                 & Total cylinder height \\
$R$                 & Cylinder radius \\
$C$                 & Circumference, $C = 2\pi R$ \\
$r_p$               & Peg radius \\
$r_b$               & Ball radius \\
$H_{\text{module}}$ & Module height (e.g.\ $8h$ for 8 peg rows) \\
$H_{\text{bin}}$    & Bin height (e.g.\ $2H_{\text{module}}$) \\
\hline
\end{tabular}
\end{table}

The angular and vertical coordinates of the pegs are
\[
\theta_{i,j} = \left(j - \frac{i}{2}\right)\Delta\theta \pmod{2\pi},
\qquad
z_i = H - i h,
\]
for $j = 0,1,\dots,i_M-1$ and $i = 0,1,\dots,n-1$. The corresponding Cartesian
coordinates are
\[
x_{i,j} = R\cos\theta_{i,j}, \qquad
y_{i,j} = R\sin\theta_{i,j}, \qquad
z_{i,j} = z_i.
\]

\begin{figure}[h]
\centering
\begin{tikzpicture}[scale=1.2]

\def\d{1.0}
\def\h{0.9}

\foreach \i in {0,1,2,3} {

    \pgfmathsetmacro{\z}{-\i*\h}

    \node[left=2pt] at (-1.6,\z+0.12) {$i=\i$};

    \foreach \j in {0,...,\i} {

        \pgfmathsetmacro{\theta}{(\j - \i/2)*\d}

        \fill (\theta,\z) circle (2pt);

        \ifnum\i=2
            \node[above] at (\theta,\z) {\scriptsize $j=\j$};
        \fi
    }
}

\node[rotate=90] at (-2.8,-1.2) {Vertical coordinate};

\end{tikzpicture}
\caption{Illustrative peg lattice showing row index $i$ and column index $j$ for
the angular coordinate $\theta_{i,j}$.}
\label{fig:lattice}
\end{figure}
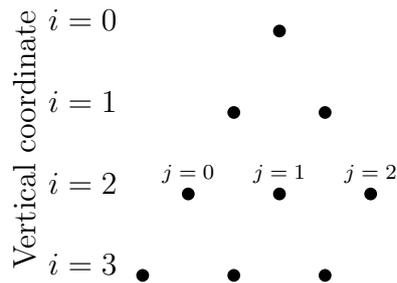

An illustrative unwrapped representation of the peg lattice, showing the row
index $i$ and column index $j$, is given in Figure~\ref{fig:lattice}.

\begin{figure}[h!]
\centering
\includegraphics[width=0.5\linewidth]{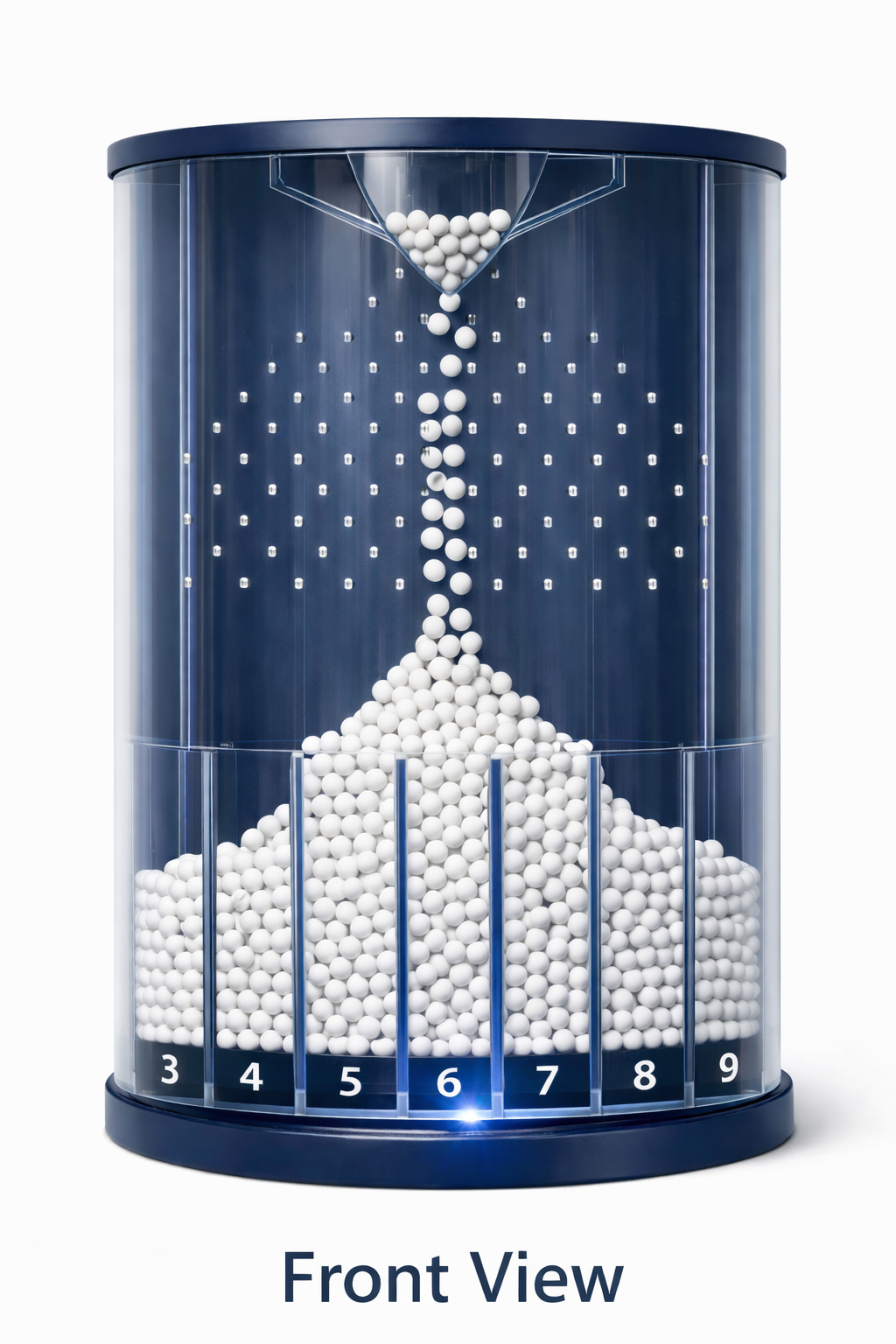}
\caption{Front view of the cylindrical Galton board obtained by wrapping a
synthetic planar Galton board. Further details are provided in  Appendix \ref{Sec: Appendix} .}
\label{fig:Frontview}
\end{figure}

Figure~\ref{fig:Frontview} shows a front view of the cylindrical Galton board,
illustrating the perspective obtained after wrapping a synthetic planar model.

\section{Wrapped Binomial and Wrapped Normal Distributions}
\label{Sec: wrapped normal}

\subsection{Wrapped binomial distribution}

We follow Section~3.5.7 of \citet{mardiajupp2000} in introducing wrapped
distributions. Let $X \sim \mathrm{Bin}(n,p)$ with support
$\{0,1,\ldots,n\}$, and let $M \ge 1$ be an integer. The \emph{wrapped
binomial distribution} on the finite cyclic group
$\mathbb{Z}_M = \{0,1,\ldots,M-1\}$ is defined by
\[
Y = X \bmod M.
\]
For $k \in \mathbb{Z}_M$, the probability mass function is
\begin{equation}\label{wBD}
 \Pr(Y=k)
   = \sum_{\ell \ge 0:\, k+\ell M \le n}
      \binom{n}{\,k+\ell M\,} p^{\,k+\ell M}(1-p)^{\,n-k-\ell M}.
\end{equation}

The characteristic function of the wrapped binomial is
\[
\phi_{\theta}(t)
= \mathbb{E}\!\left[e^{\,it\Theta}\right]
= \left(1 - p + p\, e^{\, i t 2\pi/M}\right)^{n}.
\]
For $p = 1/2$, this simplifies to
\begin{equation}\label{cfBin}
 \phi_{\theta}(t)
 = \left(\cos \frac{\pi t n}{M}\right)^{n}
   \exp\!\left(\frac{i\pi t n}{M}\right),
\end{equation}
and the first trigonometric moments are
\[
\alpha_1 = \cos^2\!\left(\frac{\pi n}{M}\right),
\qquad
\beta_1 = \cos\!\left(\frac{\pi n}{M}\right)
          \sin\!\left(\frac{\pi n}{M}\right).
\]
Hence the resultant length $\rho$ and mean direction $\mu$ are
\begin{equation}\label{1stMomBin}
\rho = \left|\cos\frac{\pi n}{M}\right|,
\qquad
\mu = \frac{\pi n}{M}.
\end{equation}

As $M \to \infty$ for fixed $n$, we have $\rho \to 1$, showing that the
wrapped binomial becomes highly concentrated at zero mean. If instead
$M \to \infty$ and $n \to \infty$ with $\delta = n/M$ fixed, then from
\eqref{cfBin} the characteristic function tends to that of the uniform
distribution on $\mathbb{Z}_M$, since
$|\cos(\pi t n/M)| < 1$ while $\exp(i\pi t n/M) = \exp(i\pi t \delta)$
remains fixed.

\begin{figure}[h!]
\centering
\includegraphics[width=0.9\linewidth]{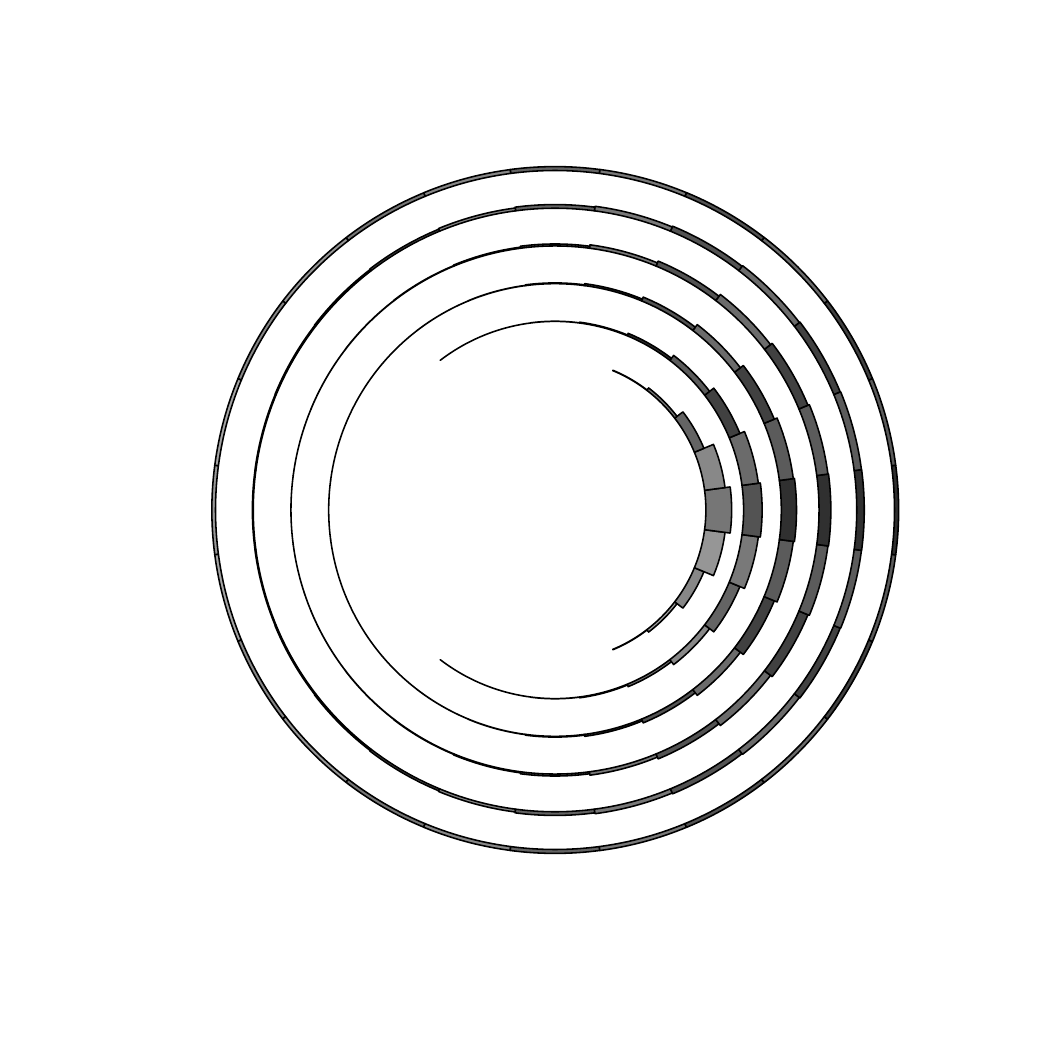}
\caption{Concentric curved barplots of wrapped binomial probabilities for
$M=24$ and $n=8,16,24,40,100,400$. The height of each bar is measured
from the corresponding circle. The support is strictly smaller than
$(-\pi,\pi]$ for small $n$, and the wrapped binomial approaches the
discrete uniform distribution for large $n$.}
\label{fig:wrapped-binomial-probabilies}
\end{figure}

For the wrapped binomial \eqref{wBD}, we compute probabilities for
various values of $n$ with $M = 24$; this choice of $M$ corresponds to
our physical model in Section~\ref{Sec: Physical}.  
Figure~\ref{fig:wrapped-binomial-probabilies} shows that $n \ge 24$ is
required for full support on $(-\pi,\pi]$, and that the distribution
approaches the discrete uniform as $n$ increases.

\subsection{Wrapped normal distribution}

The wrapped normal distribution $\mathrm{WN}(\mu,\sigma^{2})$ is defined
by wrapping a normal random variable $X \sim N(\mu,\sigma^{2})$ onto the
circle:
\[
\Theta = X \bmod 2\pi.
\]
Its probability density function on $[0,2\pi)$ is
\begin{equation}\label{WN}
  f_{\mathrm{WN}}(\theta \mid \mu,\sigma^{2})
   = \frac{1}{\sqrt{2\pi\sigma^{2}}}
     \sum_{k=-\infty}^{\infty}
       \exp\!\left(
         -\frac{(\theta - \mu + 2\pi k)^{2}}{2\sigma^{2}}
       \right).
\end{equation}

\begin{figure}[h!]
\centering
\includegraphics[width=0.7\linewidth]{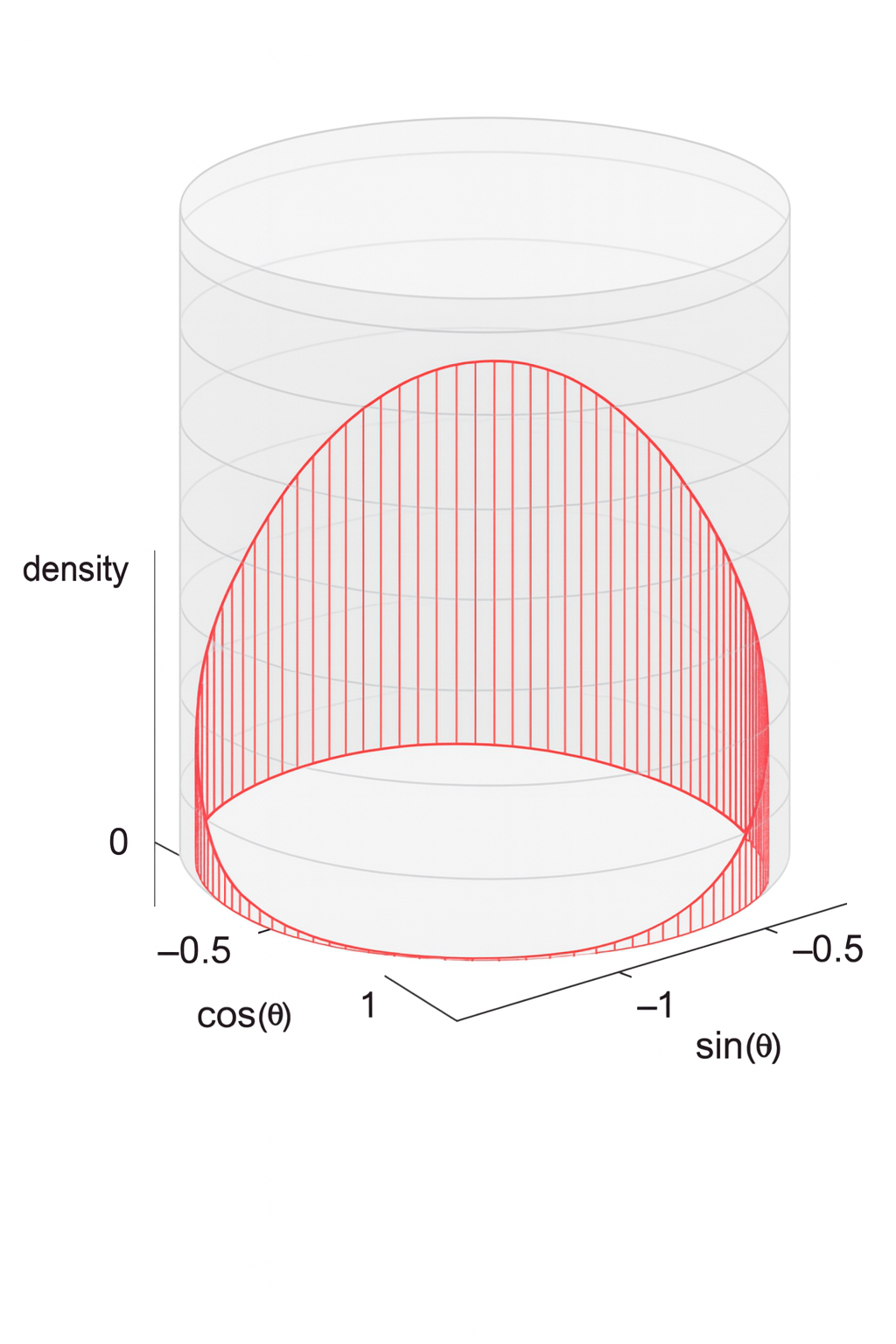}
\caption{Wrapped normal distribution plotted on the surface of a unit
cylinder. For angular coordinate $\theta\in[0,2\pi)$, the horizontal
position is $(\cos\theta,\sin\theta)$, while the vertical axis
represents the height $f_{\mathrm{WN}}(\theta)$. The density is shown
for $\mu=0$ and scale parameter $\sigma$. Vertical line segments indicate
the height above the base circle, forming a smooth crown-like surface.}
\label{fig:wrapped-normal-cylinder}
\end{figure}

Figure~\ref{fig:wrapped-normal-cylinder} illustrates a wrapped normal
distribution plotted on the interior surface of a vertical cylinder of
unit radius. The angular coordinate $\theta \in [0,2\pi)$ determines the
horizontal position $(\cos\theta,\sin\theta)$, while the vertical axis
represents the radial height of the density.

Figures~\ref{fig:Frontview},
\ref{fig:wrapped-binomial-probabilies} and
\ref{fig:wrapped-normal-cylinder} together illustrate how geometry
transforms the limiting form of distributions. On the planar domain
$\mathbb{R}$, horizontal displacement follows a standard normal law; when
wrapped onto $\mathbb{S}^1$, this becomes a wrapped normal; and when
implemented physically on a cylindrical surface, the apparatus enforces
periodic boundary conditions.

The limiting Gaussian form on $\mathbb{R}$ is consistent with classical
pin-board constructions and central limit behaviour
\citep{Galton1889, Kendall1948}. Since the planar
Gaussian lives on $\mathbb{R}$ and the cylindrical surface lives on
$\mathbb{R}/2\pi\mathbb{Z}$, projection to the periodic space induces
wrapping \citep{mardiajupp2000}, confirming the theoretical convergence.

{\bf Unimodality and the mode.}
The wrapped normal distribution is unimodal with its unique mode at the
mean direction $\mu$. Although this fact is well known, a concise proof
is not easily accessible, so we include one here. From the Fourier
series representation \citep[p.~50]{mardiajupp2000},
\[
f(\theta)
= \frac{1}{2\pi}
  \left(
    1 + 2\sum_{n=1}^{\infty}
      e^{-n^{2}\sigma^{2}/2}
      \cos\big(n(\theta-\mu)\big)
  \right),
\]
each coefficient $e^{-n^{2}\sigma^{2}/2}$ is positive for $n\ge1$. For
fixed $n$, the term $\cos(n(\theta-\mu))$ attains its maximum value $1$
at $\theta=\mu$ (mod $2\pi$) and is strictly smaller elsewhere. Since
every term in the series is maximised at $\theta=\mu$, the entire sum is
maximised at $\theta=\mu$. Uniqueness follows because at any other
$\theta$ at least one cosine term is strictly smaller than $1$. Hence
the wrapped normal has a single global maximum at $\mu$ and is unimodal.

\section{Stochastic Dynamics on the Cylindrical Board}
\label{sec:dynamics}

Given the geometric construction above, we now formalise the stochastic dynamics
of a ball descending through the cylindrical peg lattice and characterise the
resulting angular distribution at the bins.

\subsection{Ball dynamics}

Let $(\Theta_k, Z_k)$ denote the angular and vertical position of a ball after
the $k$-th deflection, with $(\Theta_0, Z_0) = (0,0)$. At each row the ball
undergoes a binary angular deflection,
\[
\Theta_{k+1}
=
\left(
\Theta_k + \xi_k \frac{\Delta\theta}{2}
\right)
\pmod{2\pi},
\qquad
Z_{k+1} = Z_k - h,
\]
where
\[
\xi_k =
\begin{cases}
+1, & \text{with probability } p,\\[4pt]
-1, & \text{with probability } 1-p,
\end{cases}
\qquad k = 1,\dots,n.
\]

Thus the ball performs a \emph{helical random walk} on the cylindrical surface.
Writing
\[
S_n = \sum_{k=1}^n \xi_k,
\qquad
\Theta_n = \left(S_n \frac{\Delta\theta}{2}\right) \pmod{2\pi},
\]
the distribution of $S_n$ is binomial,
\[
\mathbb{P}(S_n = 2k-n)
=
\binom{n}{k} p^k (1-p)^{n-k},
\qquad k = 0,\dots,n.
\]

The base of the cylindrical board is partitioned into $M$ angular slots,
\[
C_M :\quad
\theta_s = s\Delta\theta \pmod{2\pi},
\qquad s = 0,\dots,M-1,
\]
with $\Delta\theta = 2\pi/M$. The induced distribution on $C_M$ is precisely the
\emph{wrapped binomial distribution} introduced in
Section~\ref{Sec: wrapped normal}.

\subsection{Moments, drift, and asymptotics}

The mean and variance of the unwrapped angular displacement are
\[
\mathbb{E}[\Theta_n]
=
n(2p-1)\frac{\Delta\theta}{2},
\qquad
\operatorname{Var}(\Theta_n)
=
np(1-p)\Delta\theta^2.
\]
A bias $p \neq \tfrac12$ produces a systematic rotational drift.

For fixed $M$, as $n \to \infty$ the wrapped binomial distribution converges to
the discrete uniform distribution on $C_M$. This behaviour is observed both
computationally and physically: once $n > M$, balls routinely traverse past the
antipode.

For the joint limit $n \to \infty$ and $M \to \infty$ with $n/M \to \infty$, the
unwrapped displacement satisfies the classical central limit theorem,
\[
S_n \frac{\Delta\theta}{2}
\;\xrightarrow{d}\;
\mathcal{N}\!\left(
n(2p-1)\frac{\Delta\theta}{2},
\;np(1-p)\Delta\theta^2
\right),
\]
and wrapping modulo $2\pi$ yields the \emph{wrapped normal limit}
\[
\Theta_n \xrightarrow{d}
\mathcal{WN}\!\left(
\mu = n(2p-1)\frac{\Delta\theta}{2},
\;\sigma^2 = np(1-p)\Delta\theta^2
\right).
\]

If, for illustration, $n = M^\alpha$ with $\alpha > 1$, the wrapped binomial
converges simultaneously to the wrapped normal and to the uniform distribution,
so the wrapped normal limit is uniform. Conversely, when $n = M$ there is no
wrapping.

\subsection{Connection to bins}

The bins partition the circle into intervals
\[
I_k =
\left[
\frac{2\pi(k-1)}{M},
\frac{2\pi k}{M}
\right),
\qquad k = 1,\dots,M.
\]
The probability of landing in bin $k$ is
\[
p_k = \int_{I_k} f_{\mathrm{WN}}(\theta)\, d\theta,
\]
where $f_{\mathrm{WN}}$ is the wrapped normal density from the asymptotic
result above. This discrete distribution constitutes the observable output of
the cylindrical Galton board.

\subsection{Unified framework}

The transition kernel for the angular walk has the form
\[
p_{n+1}(x)
=
p\,p_n(x-1) + q\,p_n(x+1),
\]
mirroring the classical planar Galton board but on the compact group $C_M$ or
$S^1$. This identifies the wrapped normal as the natural circular analogue of
the Gaussian limit, consistent with general theory on random walks on compact
groups and circular statistics
\citep{DhunganaFragoso2016, mardiajupp2000, kato2010circular, su1998random}.

For comparison, the planar Galton board produces horizontal displacements
\[
X_n = 2(X-np)=\sum_{i=1}^n S_i,
\qquad
S_i \in \{-1,+1\},
\]
with
\[
X_n \Rightarrow \mathcal{N}\!\bigl((p-q)n,\;4npq\bigr),
\]
giving rise to the familiar near-normal distribution observed at the base of the
classical device. For the symmetrical case,
$X_n \Rightarrow \mathcal{N}\!\bigl(0,\;n\bigr).$

\section{Physical Cylindrical Model} \label{Sec: Physical}

We have constructed a physical cylindrical Galton board comprising four main components:

\begin{tabular}{lp{5in}}
\textbf{Funnel} &
A ball hopper and funnel at the top, containing approximately 1000 small steel ball bearings. \\[4pt]

\textbf{Modules} &
Five modules, each with 8 rows of pegs. Each row contains up to $M=24$ pegs with constant angular spacing 
$\Delta\theta = 2\pi/24 = 15^\circ$. Modules 1--3 contain the triangular peg lattice, while modules 4--5 
contain a rectangular lattice; all rows are staggered. \\[4pt]

\textbf{Bins} &
A set of $M=24$ equally spaced collection bins arranged around the base of the cylinder. \\[4pt]

\textbf{Frame} &
A reset mechanism that returns balls from the collection bins to the hopper by rotating the apparatus. \\
\end{tabular}

\medskip

When balls are dropped, the angular distribution after $n$ rows of pegs corresponds to a wrapped binomial 
distribution $\mathrm{WB}(n,M)$, approximating a wrapped normal (WN) distribution. Each additional row of 
pegs allows a ball to travel an additional angular displacement of $\Delta\theta/2 = 7.5^\circ$ beyond the 
antipode; we refer to this as ``wrapping to''.

\begin{table}[h]
\centering
\caption{Discrete angular distributions for the five-module physical model}
\label{tab:distributions1}
\begin{tabular}{ll}
\hline
\textbf{Modules} & \textbf{Distribution} \\
\hline
1   & Binomial with $n=8$, approximating WN on $[-60^\circ, +60^\circ]$ \\
1--2 & Binomial with $n=16$, approximating WN on $[-120^\circ, +120^\circ]$ \\
1--3 & $\mathrm{WB}(24,24)$, approximating WN with full support $(-180^\circ,180^\circ]$ \\
1--4 & $\mathrm{WB}(32,24)$, approximating WN, wrapping to $(-180^\circ,-120^\circ]$ and $[120^\circ,180^\circ]$ \\
1--5 & $\mathrm{WB}(40,24)$, approximating WN, wrapping to $(-180^\circ,-60^\circ]$ and $[60^\circ,180^\circ]$ \\
\hline
\end{tabular}
\end{table}

Each module is embedded on the surface of a right circular cylinder of radius $11.5$\,cm (4.5\,in) and height 
$8.3$\,cm (3.25\,in), containing 8 rows of pegs. Thus the peg field height ranges from 8.3\,cm (one module, 
8 rows) to 41.5\,cm (five modules, 40 rows), and can be extended further by adding additional modules.

The $i$th row of pegs ($i=1,\dots,n=40$) contains $\min(i,24)$ pegs. Modules 1--3 contain triangular 
arrangements of 1--8, 9--16, and 17--24 pegs respectively. Modules 4--5, and any subsequent modules, each 
contain $M=24$ pegs per row. Pegs in each row are centred on the $0^\circ$ midline, and successive rows are 
staggered by $\Delta\theta/2 = 7.5^\circ$, following the classical Galton board geometry. The five modules 
together contain 684 pegs: 300 in the triangular region (modules 1--3) and 384 in the rectangular region 
(modules 4--5).

The parameters of the physical model (Table~\ref{tab:physicalparameters}) were chosen to ensure that  
(i) balls fall one at a time through the peg field without interacting,  
(ii) each ball strikes exactly one peg per row,  
(iii) balls do not deflect beyond the sides of the peg field, ensured by buffers in the first three modules, and  
(iv) the five physical configurations correspond to five wrapped binomial distributions, \ref{tab:distributions1},
 and five different 
concentration parameters of the wrapped normal limit.

A further consideration is that the balls should follow the cylindrical surface, or more precisely remain 
within a cylindrical annulus. Pegs extend 6\,mm (0.25\,in) radially
 and balls have diameter 4\,mm (0.157\,in). 
The apparatus is enclosed in a clear plastic cylinder of internal diameter 12.7\,cm (5\,in) and wall thickness 
6\,mm (0.25\,in). When the collection bins are arranged around the same 4.5\,in internal cylinder, the balls 
stack in a single layer, typically in a 3--2--3--2 pattern, so that the relative heights of the balls in the 
bins form a histogram approximating the wrapped binomial distribution and the discretised wrapped normal density.

\begin{table}[h]
\centering
\caption{Parameters of the physical cylindrical Galton board}
\label{tab:physicalparameters}
\begin{tabular}{lllll}
\hline
Symbol & Description & Count & Value (cm) & Value (in) \\
\hline
$H$            & Total board height (5 modules)        &      & 61   & 24    \\
$H_p$          & Peg field height (5 modules)          &      & 41.5 & 16.25 \\
$W$            & Peg insertion board diameter    &      & 11.4 & 4.5   \\
$W^*$          & Outer cylinder inner diameter              &      & 12.7 & 5     \\
$n$            & Number of peg rows (5 modules)        & 40   &      &       \\
$d$            & Horizontal peg spacing            &      & 1.5  & 0.589 \\
$h$            & Vertical peg spacing          &      & 1.02 & 0.4   \\
$r_{\text{peg}}$  & Peg radius                   &      & 0.1  & 0.04  \\
$r_{\text{ball}}$ & Ball radius                  &      & 0.4  & 0.157 \\
$N$            & Number of balls                 & 1000 &      &       \\
$M$            & Number of bins                  & 24   &      &       \\
$b$            & Bin width                       &      & 1.4  & 0.549 \\
$H_b$          & Bin height                      &      & 13   & 5.1   \\
\hline
\end{tabular}
\end{table}

The number of modules can be increased arbitrarily by extending the height of the containing cylinder. With 
six modules ($n=48$ rows), the wrapped binomial distribution covers the circle $C_{24}$ exactly twice. 
Table~\ref{tab:distributions2} extends Table~\ref{tab:distributions1} to additional modules
Note that the division of 
rows into modules of eight is a constructional convenience rather than an intrinsic feature of the cylindrical 
quincunx.

\begin{table}[h]
\centering
\caption{Further discrete angular distributions for extended models}
\label{tab:distributions2}
\begin{tabular}{ll}
\hline
\textbf{Modules} & \textbf{Distribution} \\
\hline
1--3  & $\mathrm{WB}(24,24)$, approx WN with full support $(-180^\circ,180^\circ]$ \\
1--6  & $\mathrm{WB}(48,24)$, approx WN, wrapping once around the circle \\
1--9  & $\mathrm{WB}(72,24)$, approx WN, wrapping twice around the circle \\
1--12 & $\mathrm{WB}(96,24)$, approx WN, wrapping three times around the circle \\
\hline
\end{tabular}
\end{table}

The classical planar Galton board yields a normal distribution via the central limit theorem. When wrapped 
onto a cylinder, the Galton board induces a random walk on $\mathbb{Z} \times S^1$, producing wrapped binomial 
and wrapped normal laws governed by circular statistics.

\begin{figure}[h!]
\centering
\includegraphics[width=1.05\linewidth]{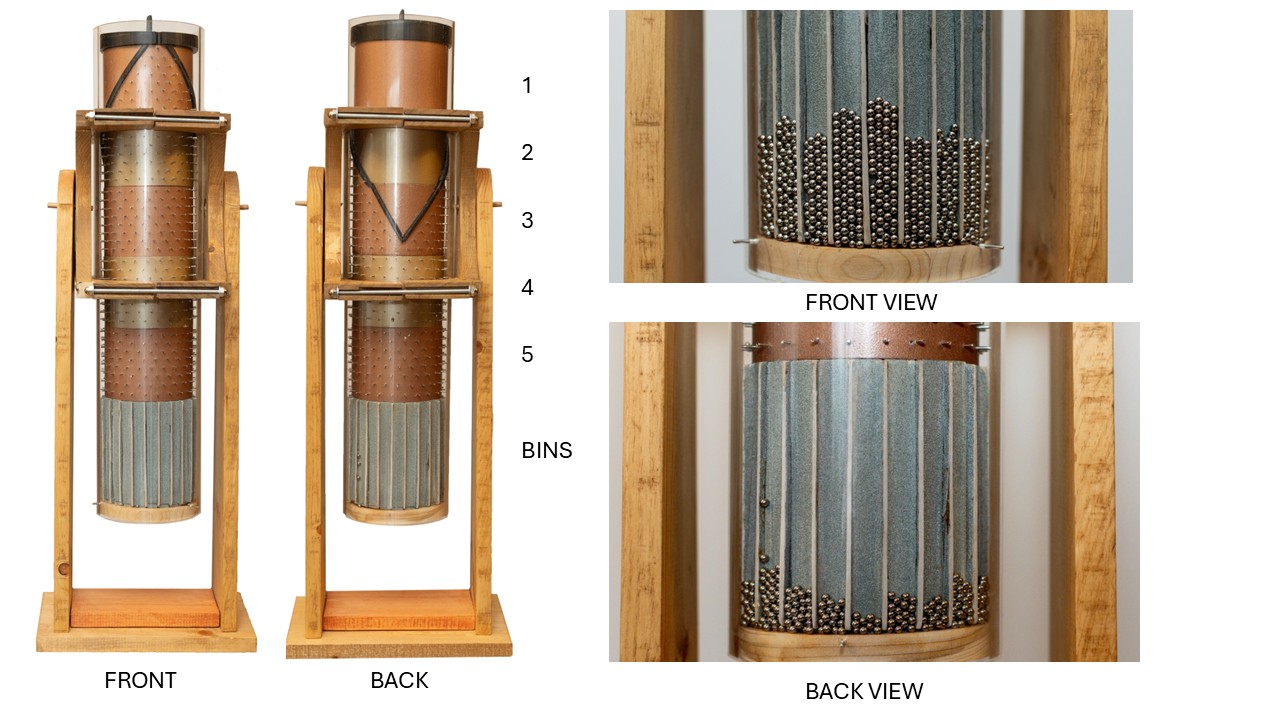}
\caption{Our physical model; left: The model showing full front and back view with Modules 1-5 and the Bins;
right: distribution of 2000 balls in the bins, front view (top) and back view (lower).}
\label{fig:OurModel}
\end{figure}
Figure~\ref{fig:OurModel} presents several views of our physical cylindrical Galton board. 
The two photographs of the apparatus on the left shows Modules~1--5 together with the collection bins,
before the balls have been  passed through the peg lattice.
The two views on the right show the distribution of 2000 balls in the collection bins,
the front view displays the modal region of the resulting 
approximate wrapped normal distribution, and the lower view shows, at the back of the device, the
the tail region of the distribution, including the impact of wrapping. 
These views demonstrate that 
the physical model generates an angular distribution that closely approximates the wrapped normal law.

\section{Discussion}\label{Sec: Disc}

The cylindrical Galton board introduced here removes the edge effects inherent in planar designs and provides a natural physical realisation of a binomial random walk with periodic boundary conditions. By enforcing angular periodicity, the device transforms the classical left--right deflection process into a random walk on the compact manifold $S^1$, thereby linking a familiar nineteenth-century apparatus to modern developments in circular statistics and stochastic processes on manifolds.

The geometry of the device plays a decisive role in shaping the resulting distributional behaviour. On the planar board, the horizontal displacement evolves on an unbounded linear domain and the classical Central Limit Theorem yields a Gaussian limit. On the cylindrical board, the same local dynamics unfold on a compact domain, producing wrapped binomial and wrapped normal laws and, for fixed $M$, eventual convergence to the discrete uniform distribution. This contrast highlights a broader principle: the topology of the state space determines the global behaviour of the random walk, even when the local transition mechanism is unchanged. The cylindrical board therefore provides a rare physical demonstration of how geometric constraints and boundary conditions influence probabilistic limits.

From a pedagogical perspective, the cylindrical Galton board offers several advantages. It provides a tangible illustration of periodic boundary conditions, a concept that is central in fields ranging from statistical physics to computational modelling. It also offers a concrete demonstration of wrapped distributions, which are typically introduced abstractly in courses on directional statistics. The modular construction allows instructors to vary the number of rows and bins, enabling students to observe directly the transition from binomial to wrapped normal to uniform behaviour. This flexibility makes the device a valuable teaching tool for illustrating the interplay between geometry, randomness, and limiting distributions.

The physical model also has methodological value. Because the device can be extended by adding modules, it provides a controlled environment for studying how the parameters $n$ and $M$ influence the emergence of wrapped distributions. This makes it suitable for simulation studies in which symmetry, invariance, and boundary behaviour are of interest. The cylindrical board can serve as a physical analogue for random walks on compact groups, offering an experimental platform for validating theoretical results in circular statistics, stochastic processes, and harmonic analysis on $S^1$.

Beyond its immediate applications, the cylindrical Galton board suggests several avenues for further investigation. One direction concerns the behaviour of biased walks, where the interaction between drift and periodicity leads to non-trivial stationary distributions. Another concerns the extension to higher-dimensional manifolds, such as random walks on tori or spheres, where physical analogues may be constructed by embedding peg lattices on curved surfaces. A third direction involves  further study of mixing times and rates of convergence to the wrapped normal or uniform distributions, which may depend sensitively on the ratio $n/M$ and on the geometry of the peg lattice.

Finally, the cylindrical Galton board offers a modern reinterpretation of Galton’s original device. Whereas Galton used the planar board to illustrate regression, variation, and the emergence of the normal distribution, the cylindrical version extends these ideas to the circular setting, connecting classical probability demonstrations with contemporary statistical theory. In doing so, it provides a bridge between historical intuition and modern geometric perspectives, illustrating how simple physical constructions can illuminate deep probabilistic principles.


\section{Acknowledgments}
We wish to thank John Kent for his helpful comments. 
\bibliographystyle{rss}
\bibliography{ref_PDWC}
\appendix
\section{Appendix: Synthetic  Cylindrical Galton Board}\label{Sec: Appendix}

\begin{figure}[h!]
  \centering
  \includegraphics[width=0.65\textwidth]{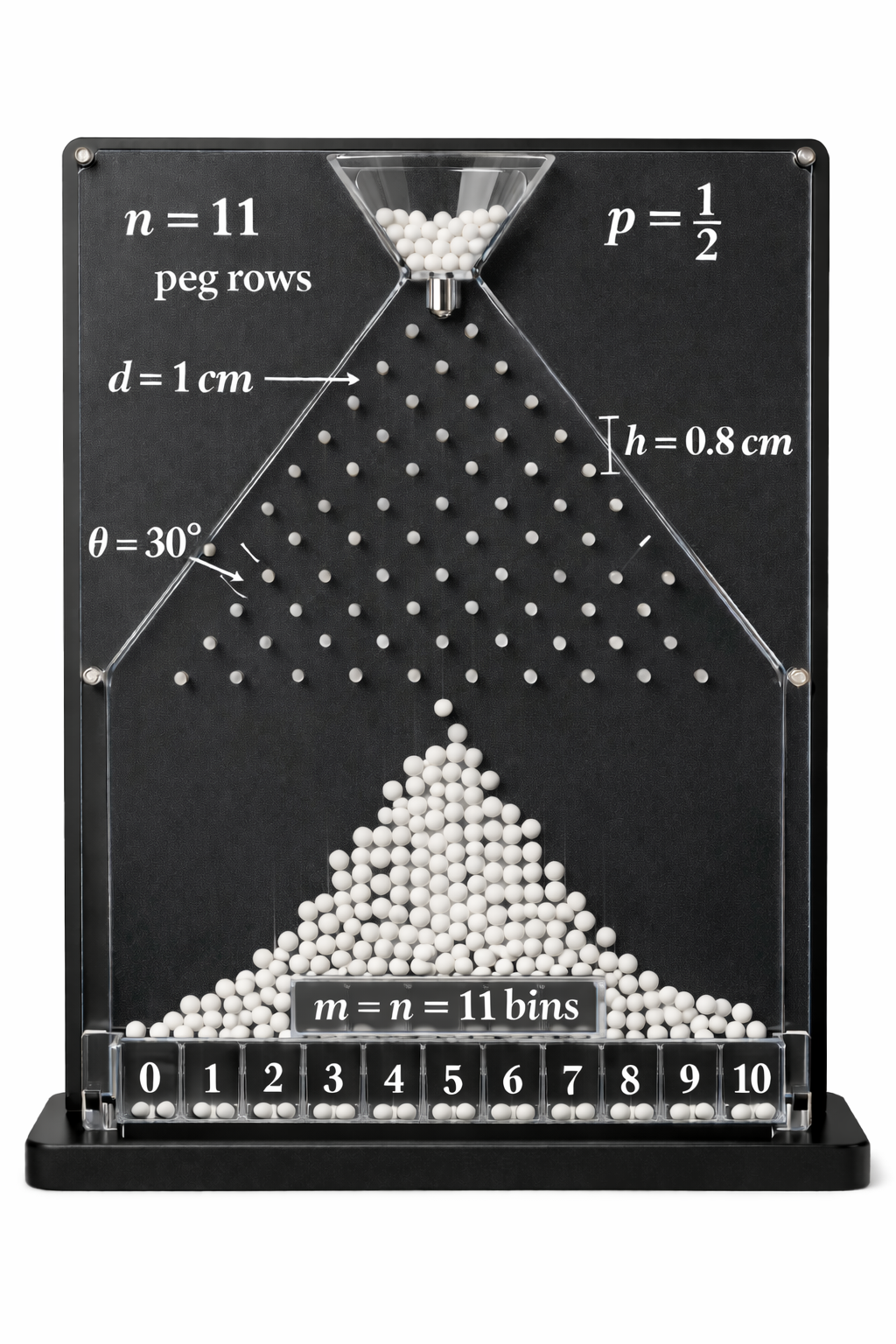}
  \caption{%
  A synthetic Galton Board Classic planar Galton board  Board size= A4 size. m=number of bins here and n= number of rows. }
  \label{fig:galton-board}
\end{figure}

We now construct a  synthetic  planar Galton board  which then wrapped onto a cylinder to create a cylindrical Galton board.  The board consists of $n=10$ rows of pegs with horizontal spacing $d=1\,\mathrm{cm}$
  and vertical spacing $h=0.8\,\mathrm{cm}$.
  Balls again undergo independent left--right deflections with probability $p=0.5$
  at each peg, resulting in $M=n+1=11$ collection bins at the base.
  The side walls are inclined at angle $\theta=30^\circ$ .
  The resulting bin counts approximate a binomial distribution and, for large $n$,
  a normal distribution as seen in  Figure \ref{fig:galton-board}.

\begin{figure}[h!]
\centering
\includegraphics[width=0.75\textwidth]{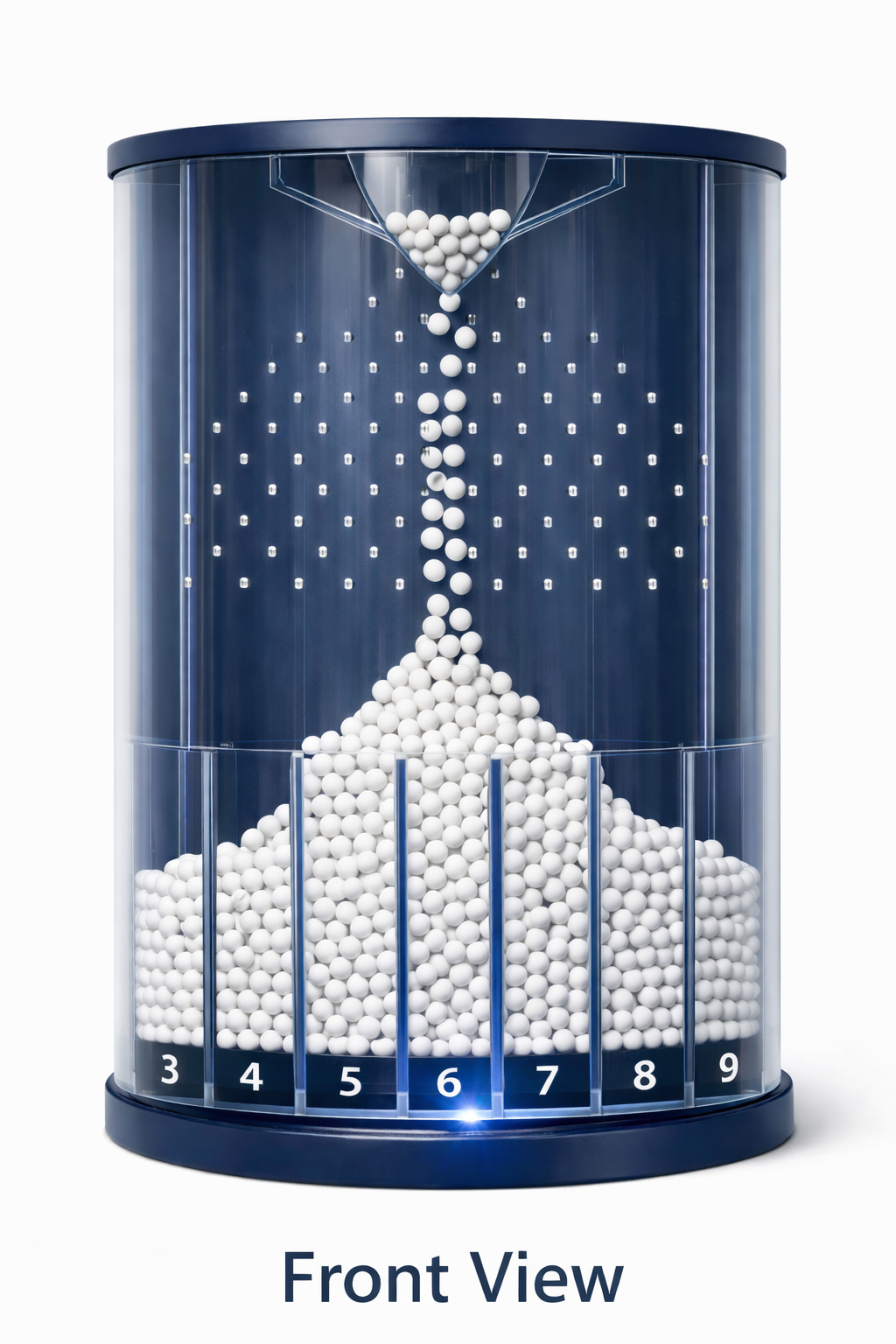}
\caption{Wrapping the synthetic Galton Board Figure  \ref{fig:galton-board}. The Front view of the cylindrical Galton board formed by wrapping the short side  of this planar A4 board. Reproduced in the Paper.}
\label{fig: synth cyl front}
\end{figure}

{\bf Wrapping the Synthetic Board.}
We now wrap our synthetic  Galton Board in Figure \ref{fig:galton-board}.
The frontal view of the wrapped cylindrical Galton board given in Figure \ref{fig:Frontview} shows the funnel, the full triangular lattice of $n=11$ peg rows, and the visible subset of collection bins at the base. The device is obtained by wrapping the short side (210\,mm) of a planar A4 Galton board, with $n=11$ peg rows and $m=11$ bins, around a cylinder of radius $r \approx 33.4\,\mathrm{mm}$ and height 297\,mm. In this projection, only the central portion of the circular bin ring is visible, so the bins are labeled from 3 to 9, corresponding to a contiguous subset of the full sequence of 11 bins. The bell-shaped accumulation of balls in these bins reflects the approximately wrapped normal distribution generated by the random left–right deflections at each peg, while the curvature of the cylindrical casing makes clear that additional bins exist beyond the visible window.

\end{document}